# Weak invariants, temporally local equilibria and isoenergetic processes described by the Lindblad equation

CONGJIE OU [1] and SUMIYOSHI ABE [1,2,3,4]

[1] *Physics Division, College of Information Science and Engineering, Huaqiao University, Xiamen 361021, China*

[2] *ESIEA, 9 Rue Vesale, Paris 75005, France*

[3] *Department of Physical Engineering, Mie University, Mie 514-8507, Japan*

[4] *Institute of Physics, Kazan Federal University, Kazan 420008, Russia*

**Abstract** – The concept of weak invariants is examined in the thermodynamic context. Discussions are made about the temporally-local equilibrium states, corrections to them, and *isoenergetic* processes based on the quantum master equations of the Lindblad type that admit time-dependent Hamiltonians as weak invariants. The method for determining the correction presented here may be thought of as a quantum-mechanical analog of the Chapman-Enskog expansion in nonequilibrium classical statistical mechanics. Then, the theory is applied to the time-dependent harmonic oscillator as a simple example, and the power output and the work along an isoenergetic process are evaluated within the framework of finite-time quantum thermodynamics.

Recent developments in research of contemporary importance such as quantum engines with finite power outputs [1], quantum information processing under the influence of the environment [2], and biological processes including photosynthesis [3] (see also ref. [4] and the references therein) have revived interest in the physical foundations of the theory of subdynamics governing open quantum systems. Accordingly, much attention is currently being focused on the diverse physical properties of both Markovian and non-Markovian quantum subdynamics. In particular, much in non-Markovianity [5] remains to be explored and is therefore under vital investigation now in anticipation of the fact that it may provide valuable quantum resources through information backflows [6,7], for example. Even in the Markovian cases, development of the theory is also ongoing. In refs. [8,9], expansion/compression processes of the specific type termed isoenergetic processes have been studied in the context of quantum thermodynamics. These processes are defined in such a way that the internal energy of the subsystem under consideration is kept constant along them. Since the law of equipartition of energy is violated in quantum mechanics, isoenergetic processes are different from familiar isothermal processes. This, in turn, leads to a possibility of revealing a new role of the quantum effects in thermodynamics of small systems. For such processes exotic in view of traditional thermodynamics, it is necessary to introduce the environmental system of a specific type that is referred to as the *energy bath*. Recent works on energy transfer [10,11] show the possibility of implementing isoenergetic processes. It is also pointed out that the concept of conserved internal energies of subsystems can be extended to a more general one termed the weak invariants, which have recently been studied for the damped harmonic oscillator in quantum mechanics [12] and for the classical Fokker-Planck equation [13]. The notion of energy bath indicates how quantum mechanics widens the concept of the bath in classical thermodynamics, i.e., the heat bath. In this respect, another bath to be mentioned, which is also characteristic in quantum mechanics, is the dephasing bath inducing decoherence [14].

It is worth mentioning that the difference between the isothermal and isoenergetic processes indicates presence of ensemble-dependent dynamics and such a situation can also occur in open classical small systems [15].

In this paper, we discuss the physical properties of the temporally-local equilibrium states based on the Markovian quantum master equations of the Lindblad type [16,17] that describe isoenergetic processes, along which time-dependent Hamiltonians are weak invariants. The problem of determining corrections to the temporally-local equilibrium states is somewhat analogous to the Chapman-Enskog method [18], which is a standard one in nonequilibrium statistical mechanics for derivation of hydrodynamics from classical kinetic theory. As a simple example of possible applications of the theory developed here, a finite-time isoenergetic process of the time-dependent harmonic oscillator is analyzed by the use of the Lindbladian operator introduced in ref. [19], and both the power output and the work are explicitly evaluated along the process.

Let us start our discussion with succinctly explaining the concept of weak invariants [12]. Consider a general master equation

$$i\hbar \frac{\partial \rho}{\partial t} = \pounds(\rho), \tag{1}$$

where $\rho$ is the density matrix describing a quantum state of an open system, and $\pounds$ is a certain linear superoperator acting on $\rho$. Then, a weak invariant, $I(t)$, associated with eq. (1) is defined as a solution of the following equation:

$$\frac{\partial I(t)}{\partial t} - i\,\pounds^*\bigl(I(t)\bigr) = 0, \tag{2}$$

where $\pounds^*$ is the adjoint of $\pounds$, and here and hereafter $\hbar$ is set equal to unity. The eigenvalues of $I(t)$ are not constant in time, in general, but the expectation value is:

$$\frac{d}{dt}\,\mathrm{tr}\bigl[I(t)\rho\bigr] = 0, \tag{3}$$

where $\rho$ is a solution of eq. (1) satisfying an arbitrary initial condition. $I(t)$ is a generalization of the Lewis-Riesenfeld invariant [20] of a system governed by the Schrödinger equation with a time-dependent Hamiltonian and is referred to as a strong invariant.

A primary difference between strong and weak invariants is that the spectrum of the former is constant in time, whereas in the latter it is not, in general, as mentioned above [12].

If subdynamics is Markovian, then the most general master equation preserving positive semidefiniteness of a density matrix is known to be uniquely of the form [16,

17]:

$$i\frac{\partial \rho}{\partial t} = [H(t), \rho] - i\sum_n c_n \left( L_n^\dagger L_n \rho + \rho L_n^\dagger L_n - 2 L_n \rho L_n^\dagger \right).  \quad (4)$$

Here, $H(t)$ is the Hamiltonian of a subsystem under consideration and is allowed to have explicit time dependence. $L_n$'s are the so-called Lindbladian operators, and $c_n$'s are *c*-number coefficients that have to be nonnegative in order for the density matrix to remain positive semidefinite in the course of time evolution. Both the Lindbladian operators and the coefficients, too, may generically depend on time. Then, the weak invariant, $I(t)$, associated with eq. (4) is a solution of the following equation [12,21,22]:

$$\frac{\partial I(t)}{\partial t} + i[H(t), I(t)] - \sum_n c_n \left[ L_n^\dagger L_n I(t) + I(t) L_n^\dagger L_n - 2 L_n^\dagger I(t) L_n \right] = 0.  \quad (5)$$

If all of $c_n$'s vanish, then $I(t)$ becomes reduced to the Lewis-Riesenfeld strong invariant of the time-dependent Hamiltonian [20]. Later, we shall discuss a weak invariant given by the Hamiltonian itself and its role in finite-time quantum thermodynamics.

To specify the Lindbladian operators and the *c*-number coefficients, details have to be known about the interaction between the subsystem and the environment, in general. However, in a particular case where the time-dependent Hamiltonian is required to be a weak invariant, the situation can drastically be simplified [19], as will be seen below.

Among various quantum states of the subsystem with the time-dependent Hamiltonian, of particular interest is the temporally-local equilibrium state, which is written as follows:

$$\rho^{(0)} = \frac{1}{Z(\beta(t))} e^{-\beta(t) H(t)},  \qquad (6)$$

where $\beta(t)$ is the temporally-local inverse temperature, and $Z(\beta(t))$ is the partition function given by

$$Z(\beta(t)) = \text{tr } e^{-\beta(t) H(t)}. \qquad (7)$$

The temporally local equilibrium state in eq. (6) is analogous to the local equilibrium Maxwell-Boltzmann distribution in nonequilibrium classical statistical mechanics [18], around which the Chapman-Enskog expansion of the classical Boltzmann equation is performed in the relaxation-time and hydrodynamic approximations. There, the temperature, the average velocity and the particle number density depend on both space and time coordinates and satisfy the equations in fluid dynamics. In the present case, on the other hand, the (inverse) temperature is a function only of time, since it is not a field and should not depend on the position operator.

The state in eq. (6) may be thermodynamically relevant if the time scale of change of the Hamiltonian is much larger than the quantum dynamical one, $\sim 1/E$ (i.e., $\sim \hbar/E$) with $E$ being a typical value of the energy. Change of the local inverse temperature in time should also be slow. If the coefficients, $c_n$'s, are small and are of the same order as those of the time derivatives of the Hamiltonian and the inverse temperature, then eq. (4) enables us to systematically determine corrections to the temporally local equilibrium state in eq. (6). To see this, let us write the density matrix as follows:

$$\rho = \rho^{(0)} + \rho^{(1)} + \cdots. \tag{8}$$

The correction terms have to be traceless, since both $\rho$ and $\rho^{(0)}$ are normalized. Substitution of eq. (8) into eq. (4) leads, in the first order, to

$$[H(t), \rho^{(1)}] = iR, \tag{9}$$

where

$$R \equiv \dot{\rho}^{(0)} + \sum_n c_n \left( L_n^\dagger L_n \rho^{(0)} + \rho^{(0)} L_n^\dagger L_n - 2 L_n \rho^{(0)} L_n^\dagger \right). \tag{10}$$

$\dot{\rho}^{(0)}$ in eq. (10) is understood as the time derivative through the changes of the time-dependent parameters contained. This is also analogous to the Chapman-Enskog expansion method [18]. The formal solution of eq. (9) can be obtained by the use of the formula

$$\rho_\varepsilon^{(1)} = \int_0^\infty d\lambda \, e^{i\lambda H(t)} R \, e^{-i\lambda H(t) - \varepsilon \lambda}, \tag{11}$$

where $\varepsilon$ is a positive infinitesimal constant that makes the integral convergent. The equation satisfied by $\rho_\varepsilon^{(1)}$ is: $[H(t), \rho_\varepsilon^{(1)}] = iR - i\varepsilon \rho_\varepsilon^{(1)}$. The limit $\varepsilon \to 0+$ has to be taken after the integration with respect to $\lambda$.

In the above discussion, there exists an important point to be noted. Clearly, eq. (9) is invariant under the transformation,

$$\rho^{(1)} \to \rho^{(1)} + \sigma, \tag{12}$$

where $\sigma$ is a certain traceless matrix that commutes with $H(t)$. Since $\rho^{(1)}$ is desired to be proportional to $\rho^{(0)}$ as in the Chapman-Enskog expansion [18], $\sigma$ has the form

$$\sigma = \left(A - \langle A \rangle_0\right)\rho^{(0)}, \tag{13}$$

which is manifestly traceless, provided that $A$ is a certain operator that commutes with $H(t)$ and thus also with $\rho^{(0)}$, and the symbol $\langle \bullet \rangle_0$ stands for the expectation value with respect to the temporally local equilibrium state, that is,

$$\langle Q \rangle_0 \equiv \text{tr}\left(Q \rho^{(0)}\right). \tag{14}$$

This symmetry may remind us of the gauge-theoretic structure in thermodynamics discussed in ref. [23]. Therefore, any physical (e.g., thermodynamic) quantity should be independent of this arbitrariness.

Now, let us consider as a simple example the time-dependent harmonic oscillator.

The Hamiltonian reads

$$H(t) = \frac{1}{2}p^2 + \frac{1}{2}k(t)x^2, \qquad (15)$$

where $k(t)\,(>0)$ is the time-dependent spring coefficient. This model may be regarded as a simplified system of a particle in a time-dependent harmonic trap, for example. It has been shown [19] that the Lindblad equation can admit the Hamiltonian in eq. (15) as a weak invariant only with a single coefficient and a corresponding single Lindbladian operator,

$$c_1 = -\frac{\dot{k}(t)}{8}, \qquad (16)$$

$$L_1 = x^2, \qquad (17)$$

respectively. Since the coefficient, $c_1$, should be positive, the harmonic potential has to be widening:

$$\dot{k}(t) < 0. \qquad (18)$$

In other words, as long as based on the Lindblad equation, only expansion is allowed as the isoenergetic process. Thus, *the concept of weak invariants automatically determines both the coefficients and the Lindbladian operators without detailed knowledge about the interaction between the subsystem and the environment and puts a stringent* *constraint on the processes generated by the Lindblad equation.* Accordingly, eq. (4) becomes

$$i\frac{\partial \rho}{\partial t}=[H(t),\rho]+\frac{i\dot{k}(t)}{8}[x^2,[x^2,\rho]].\qquad(19)$$

In contrast to the work in ref. [24], the operator appearing in the double commutator is not linear in *x* but quadratic. $\dot{k}(t)$ plays a role of a small parameter in the theory. Since the potential is widening, the energy spectrum is shifting lower. Hence, it is necessary for the subsystem to receive energy in order to preserve the internal energy, i.e., the expectation value of the Hamiltonian. That is, energy transfer to the subsystem from the outside is necessary. Recent developments [10,11] seem to suggest that it may, in fact, be realized. A process, along which the internal energy is kept constant, is refereed to as an isoenergetic process [8,9]. Accordingly, "the outside" system mentioned above is called the energy bath [9,25], which is usually unnecessary in classical thermodynamics since there the energy transfer between the subsystem (*e.g.*, an engine) and the environment is heat exchange. As mentioned at the very beginning of this paper, such a process emphasizes how quantum thermodynamics can be different from classical thermodynamics due to the quantum-mechanical violation of the law of equipartition of energy.

Since the time-dependent Hamiltonian in eq. (15) is a weak invariant associated with the Lindblad equation in eq. (19), an isoenergetic process can be treated within the framework of finite-time quantum thermodynamics. In what follows, we discuss this issue in detail.

The partition function appearing in the temporally-local equilibrium state is a familiar one: $Z(\beta(t)) = 1/\{2\sinh[\beta(t)\sqrt{k(t)}/2]\}$. (Recall that, in terms of the frequency, $\omega$, $\sqrt{k} = \hbar\omega$ with $\hbar = 1$ and the unit mass.) Since the time dependence of the inverse temperature comes from that of $k(t)$ contained in the Hamiltonian, $\dot{\beta}(t)$ is of the order of $\dot{k}(t)$, which is a small parameter in the theory, as can be seen in eq. (21) below. Therefore, it is necessary to find the relation between $\beta(t)$ and $k(t)$ to characterize the isoenergeticity condition. Calculation of $\rho_\varepsilon^{(1)}$ in eq. (11) with

$$R = \dot{\rho}^{(0)} - \frac{\dot{k}(t)}{8}[x^2,[x^2,\rho^{(0)}]] \tag{20}$$

and $\dot{\rho}^{(0)} = \dot{k}(t)\,\partial\rho^{(0)}/\partial k + \dot{\beta}(t)\,\partial\rho^{(0)}/\partial\beta$ is much involved, and its long expression is not presented, here. As will be seen below, the explicit form of $\rho^{(1)}$ is not needed for the subsequent discussion in the lowest order.

According to eq. (8), the internal energy is also expressed as follows: $U = U^{(0)} + U^{(1)} + \cdots$, where $U^{(0)} = \langle H(t)\rangle_0 = (\sqrt{k(t)}/2)\coth[\beta(t)\sqrt{k(t)}/2]$ in the notation in eq. (14), $U^{(1)} = \text{tr}(H(t)\rho^{(1)})$ and so on. The isoenergeticity condition in the lowest order, $dU^{(0)}/dt = 0$, is required to hold. This requirement leads to the relation

$$\left(\sqrt{k(t)}\right)^{\cdot}\left[\sinh\left(\beta(t)\sqrt{k(t)}\right) - \beta(t)\sqrt{k(t)}\right] = k(t)\dot{\beta}(t). \tag{21}$$

Because of eq. (18) and the inequality $\sinh x - x > 0$ for $x > 0$, $\dot{\beta}(t)$ is negative, implying that the temporally-local temperature monotonically increases in time.

Finally, let us look at the power output and the work along an isoenergetic process. The power output, $P(t) = -\text{tr}[\partial H(t)/\partial t \, \rho]$, in the lowest order is given by

$$P(t) = -\frac{\dot{k}(t)}{2}\langle x^2 \rangle_0$$

$$= -\frac{\dot{k}(t)}{4\sqrt{k(t)}}\coth\left[\frac{1}{2}\beta(t)\sqrt{k(t)}\right]$$

$$= -\frac{\dot{k}(t)}{2k(t)}U^{(0)}. \tag{22}$$

Then, the work is obtained by the time integral of the power output

$$W = \int_{t_i}^{t_f} dt\, P(t)$$

$$= \frac{1}{2}U^{(0)}\ln\frac{k(t_i)}{k(t_f)}, \tag{23}$$

where $t_i$ ($t_f$) is the initial (final) time and the value of $U^{(0)}$ is taken to be its initial one: $U^{(0)} = \left(\sqrt{k(t_i)}/2\right)\coth\left[\beta(t_i)\sqrt{k(t_i)}/2\right]$, for example, since it is conserved. In the lowest order, $\rho^{(1)}$ is not needed for the calculations of these quantities, as mentioned earlier.

The results in eqs. (22) and (23) offer finite-time generalizations of the isoenergetic processes in quantum thermodynamics discussed in refs. [8,9], although considered there is not the harmonic potential but the infinite square well potential. It should be noted that, as long as based on the Lindblad equation, the isoenergetic compressions considered in refs. [8,9] are ruled out, as can be seen in eq. (18). Thus, preservation of positive semidefiniteness of the density matrix puts a stringent constraint on isoenergetic processes. Physically, this fact is interpreted as follows. First of all, the Lindbladian operator in eq. (17) is "normal", that is, $\left[L_1^\dagger, L_1\right] = 0$ because of self-adjointness of $L_1$ (equivalently, the dynamics is "unital"). Accordingly, the system has a positive entropy production rate [26,27], which means the irreversible dissipative nature of the dynamics. Therefore, the system is deprived of its energy. This is consistent with the expansion of the system size, since expansion lowers the energy levels. Thus, in order for the internal energy to be preserved, transfer of the energy from the energy bath to the subsystem is essential.

In conclusion, we have formulated a quantum isoenergetic processs by making use of the Lindblad equation and the time-dependent Hamiltonian as a weak invariant associated with the equation. We have applied the theory developed to the time-dependent harmonic oscillator in the temporally-local equilibrium state with the lowest order correction and have explicitly evaluated the power output and the work.

* * *

CO acknowledges the supports by the grants from Fujian Province (No. 2015J01016, No. 2016J01021, No. JA12001, No. 2014FJ-NCET-ZR04). SA has been supported in part by the grants from National Natural Science Foundation of China (No. 11775084), from Grant-in-Aid for Scientific Research from the Japan Society for the Promotion of Science (No. 16K05484), and by the Program of Competitive Growth of Kazan Federal University from the Ministry of Education and Science of the Russian Federation. This work has been completed while SA has stayed at the Wigner Research Centre for Physics with the support of the Distinguished Guest Fellowship of the Hungarian Academy of Sciences. He would like to thank the Wigner Research Centre for Physics for the warm hospitality.